# Improving Style Similarity Metrics of 3D Shapes

Kapil Dev*    Manfred Lau**

Lancaster University, UK

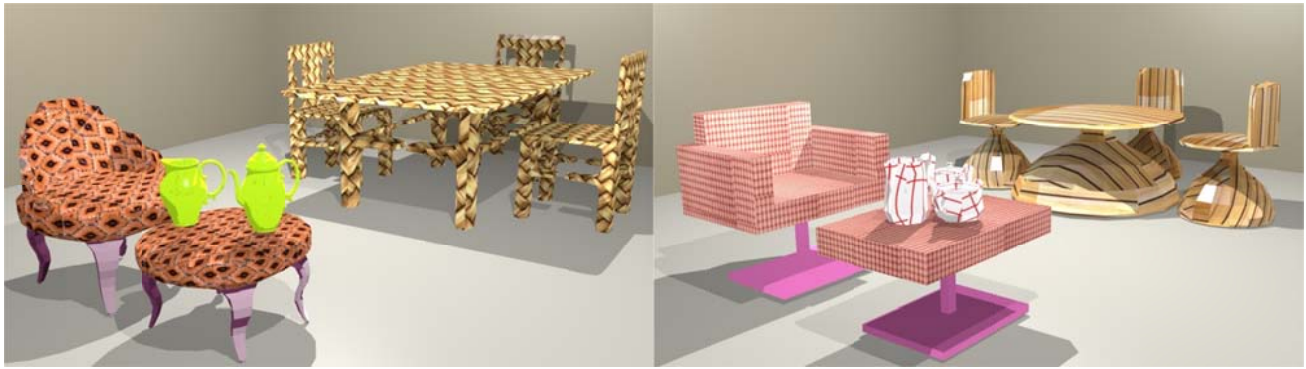

Figure 1: Two examples of 3D scene composition created with our style similarity metric that considers geometry, color, and texture.

**ABSTRACT**

The idea of style similarity metrics has been recently developed for various media types such as 2D clip art and 3D shapes. We explore this style metric problem and improve existing style similarity metrics of 3D shapes in four novel ways. First, we consider the color and texture of 3D shapes which are important properties that have not been previously considered. Second, we explore the effect of clustering a dataset of 3D models by comparing between style metrics for a single object type and style metrics that combine clusters of object types. Third, we explore the idea of user-guided learning for this problem. Fourth, we introduce an iterative approach that can learn a metric from a general set of 3D models. We demonstrate these contributions with various classes of 3D shapes and with applications such as style-based similarity search and scene composition.

**Index Terms**: I.3.5 [Computer Graphics]: Computational Geometry and Object Modeling—Modeling Packages

## 1 INTRODUCTION

Metric learning was introduced in the machine learning community for computing distance functions [Kul13]. The concept of metric learning has recently been applied to various problems in computer graphics for computing distance functions that correspond to style similarity metrics for 2D clip art [Gar14], infographics [Sal15], and 3D shapes [Liu15, Lun15]. Our work is inspired by these previous works and we focus on style similarity metrics of 3D shapes in this paper. We start with the same overall framework of computing 3D shape features, collecting human preferences of style data with crowdsourcing, and learning a style similarity distance function between pairs of 3D shapes. In contrast, we improve existing work with four novel contributions.

* k.dev@lancaster.ac.uk
** manfred.lau@gmail.com

Our first contribution is in considering the color and texture of 3D shapes in addition to their geometry within the style similarity metric. To the best of our knowledge, the state-of-the-art previous works [Liu15, Lun15] focus only on geometric features. We compute additional features corresponding to the color and texture of each 3D model, and study the role of these features towards the style metric in addition to geometric features such as curvatures, shape distributions, and shape diameter functions. We hypothesize that color and texture features would be dominant over geometric features. We test this hypothesis by observing the relative values of the set of learned weights that correspond to the features, and by observing the results of style-based similarity searches of 3D models with colors and textures.

The second contribution considers learning style metrics on a dataset of 3D models by clustering them. If the 3D shapes represent multiple object types, we can cluster them and learn style metrics in different ways. We can have a metric for each pair of object types (e.g. *chairs→tables*, *forks→spoons*) which we hypothesize will be more accurate, but there will be $N^2$ metrics if there are $N$ object types. We can also build clusters of object types (e.g. "furniture" for chairs and tables, "cutlery" for forks and spoons). We will have $K^2$ metrics if there are $K$ clusters, which can be much smaller if $K$ is much smaller than $N$. However, we hypothesize that these metrics will be less accurate as they combine 3D shapes of different types. While Liu et al. [Liu15] compared between learning from triplets (i.e. data generated from queries) of two object types (e.g $X→Y$, and $Y→X$) and all triplets, they did not explore further the clustering of object types that we propose. Our "clustering" is intuitive as the 3D models are typically divided into high-level clusters such as cutlery and then more specific object types such as forks, spoons, and knives.

The third contribution is to explore the idea of user-guided learning for the style metric problem. We experiment with learning both generic and user-guided metrics of style similarity. The generic metric is based on the crowdsourced style matching preferences, while a user-guided metric is based only on one user's style preferences (although they can also be combined). If a user is not satisfied with the search results from a crowdsourced metric, our interface allows the user to provide information (e.g. re-rank the results) and create new training data for learning a user-guided style similarity metric. We hypothesize that this user-guided concept will be beneficial in some cases.

Our fourth contribution is to introduce an iterative approach to learn a metric that can learn from a general set of 3D models. The motivation is that previous work constructs the crowdsourcing queries either randomly which can lead to many queries that provide irrelevant data [Liu15], or by manually placing all 3D models into carefully-constructed groups in order to generate useful queries [Lun15]. We thereby develop an approach where an initial set of queries are used to learn a metric, which is then iteratively used to generate further queries and metrics. We hypothesize that this iterative process can generate useful queries without tedious manual processing.

We demonstrate the above four contributions with various classes of 3D shapes (e.g. furniture, tableware, and cutlery) and build tools to show the applications of style-based similarity search and 3D scene composition. We obtain empirical results to test our hypothesis in each case. Our results will help to improve the development of style similarity metrics of 3D shapes.

### 1.1 Related Work

**Style Similarity Metrics.** There has been a recent interest in research in style-based similarity metrics of both 2D and 3D content. The idea is to compute a style-based distance function between pairs of 2D or 3D objects. In the case of 2D content, such distance functions have been developed for 2D clip art [Gar14], font selection [ODo14], and infographics [Sal15].

In the case of 3D content, the state-of-the-art methods for computing style similarity metrics of 3D shapes [Liu15, Lun15] are most closely related to our work. Liu et al. constructs part-aware feature vectors for predicting style compatibility between 3D furniture models from different object classes. Lun et al. creates a style-similarity measure based on geometric elements of the 3D shapes. Our work is different in the four contributions described above.

In particular, we give more details here on how data is collected in previous methods that is different from our approach. While Liu et al. [Liu15] learn a distance metric on randomly generated crowdsourcing queries, Lun et al. [Lun15] learn it on crowdsourcing queries that are selected from groups of 3D models that are manually pre-classified. However, if the number of 3D models is large, constructing queries (i.e. in the form of triplets of 3D models) at random can result in a large number of queries for which humans do not provide consistent responses. Such queries do not provide useful information for the learning process. On the other hand, Lun et al. [Lun15] constructs most of the queries by first manually placing the 3D models into meaningful groups, from which more useful queries can be generated. However, a manual pre-processing of the 3D models is required. In addition, these queries typically have obvious responses such that the crowdsourcing step seems unnecessary. In this paper, we thereby take an iterative approach to learn a metric, somewhat similar to an adaptive selection method to generate queries [Tam11]. Our iterative approach does not generate random queries (except possibly in the first iteration) and does not require a manual pre-grouping of the 3D models.

**3D Shape Retrieval.** Our work is related to the area of 3D shape retrieval [Fun05, Tan08] as one of our applications is in style-based search of 3D shapes. Our key difference is in the "style-based" aspect with our crowdsourced and learned style metric.

**Personalized Content Retrieval and User-Guided Learning.** Personalized information retrieval [Pas10] involves learning a user specific model of perceived relevance to present reordered search results. There exists previous work in personalized image search [San12, Kov13]. Prior work in 3D content-based retrieval focuses on learning with crowdsourced data. CueFlik [Fog08] allows users to re-rank search results for the problem of image search, and to create their own rules which can then be used to improve the search. Our work explores the idea of user-guided learning to learn user-guided style metrics for 3D shapes.

### 1.2 Overview

Figure 2 shows an overview of our approach. We collect a dataset of 3D models from online sources (step 1). We then compute various shape descriptors or features (including color/texture features) for each 3D model (step 2). We generate queries containing triplets of 3D models and place them on Amazon Mechanical Turk to collect crowdsourced data regarding style preferences of the 3D shapes (step 3). The features and collected data are then used to compute a style similarity measure with an iterative approach (step 4). The style metric can be used in various applications, including style-based search of 3D models (step 5). An individual user can re-rank the models in our interface according to their style preferences, and this information can then be used to compute a user-guided style metric.

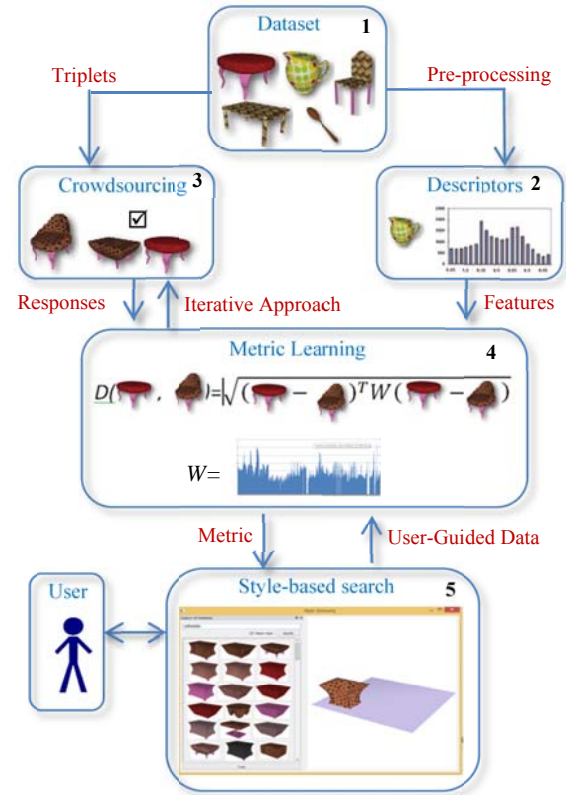

Figure 2: Overview of our framework. Our novelty is in considering color/texture features (step 2), exploring the learning of metrics with different clusters of 3D shapes (all steps), user-guided learning (steps 4 and 5), and an iterative approach (steps 3 and 4).

## 2 DATASET AND 3D MESH FEATURES

### 2.1 Dataset

We collected a dataset of 3D models (Figure 3) from the following sources: 3D Warehouse, Threeding.com, Thingiverse,

and Lun et al. [Lun15]. The object types can be categorized into "tableware" (teapots, sugar bowls, creamers), "cutlery" (knives, spoons, forks), "living" room furniture (sofas, coffee tables), and "dining" room furniture (chairs, tables). Our choice of texture images is inspired by commonly used patterns in real life. For example, dining room furniture mainly uses wooden shades and textures while living room furniture uses fabric shades and textures. We have 17 models x 5 textures for each type of tableware, 21 x 7 for each type of cutlery, 18 x 7 for each type of living room furniture, and 21 x 7 for each type of dining room furniture.

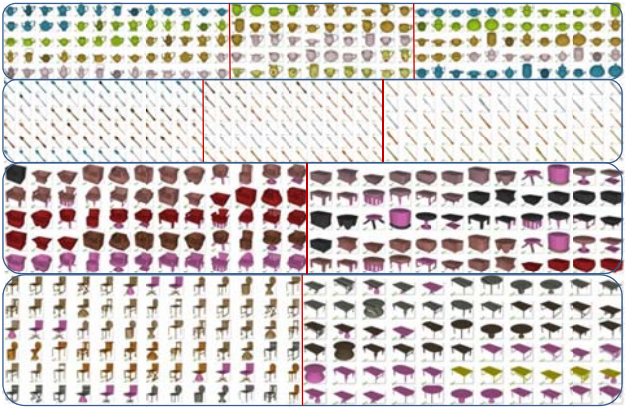

Figure 3: Example 3D models. The rows correspond to tableware, cutlery, living room furniture, and dining room furniture.

## 2.2 Geometric Features

We compute shape descriptors on the dataset of 3D models. We represent each vertex of each 3D model with a 2728-dimensional feature vector (**x**). Before computing features, all models are oriented in the same direction and scaled to have similar proportions within each object type. We use an over-complete set of features and let the learning decide the relative importance of each feature.

We aim to capture both global and local shape properties. The features are not new on their own. Please refer to previous work [Osa01, Sur03, Che03, Lun15] for details of them. We include the histograms for the following features (with the number of histogram bins in brackets): shape distribution (128), curvature (gauss, mean, max, min: 128 each), shape diameter (128), light field descriptor (470), voxel gradient (192), voxel gradient direction (128), silhouette centroid distances (192), silhouette Fourier descriptor (57), silhouette Zernike moments (108), silhouette D2 descriptor (192), silhouette gradient (192), silhouette gradient direction (96), and shape histogram (192). For these geometric features, there are a total of 2587 dimensions in the feature vector.

The first three features above are computed on a dense uniformly-sampled version of a model's surface. For the remaining features above, the model is voxelized (300×300×300), and silhouettes along x, y, and z directions are obtained by projecting the voxel space along the three axes respectively.

## 2.3 Color and Texture Features

We capture color and texture properties with the following features (inspired by [Gar14]): average HSV of the top five dominant colors (3), hue histogram (32), saturation histogram (32), value histogram (32), and local binary patterns (42). Our feature vector has a total of 141 dimensions of such features.

These features are extracted for each 3D model from its associated information about material and texture (.mtl and .png files). Figure 4 shows the textures that we used. To maintain uniformity, all texture images are cropped and resized to 512x512.

A 3D shape may have more than one texture. For example, a 3D model of a chair may consist of a wood texture for its seating area and a steel texture for its legs. We handle multiple textures by computing the same features above for each texture and combining their histograms.

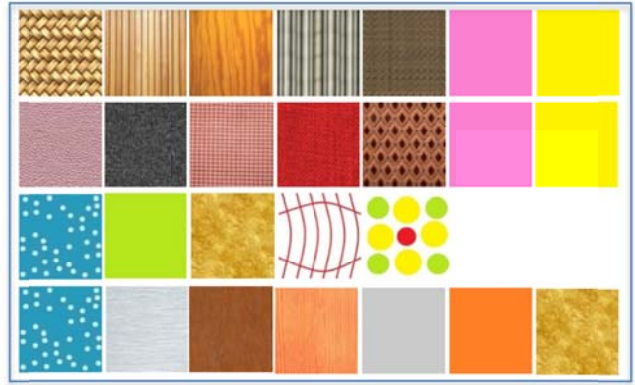

Figure 4: Textures used with our 3D models. The rows correspond to "dining", "living", "tableware", and "cutlery".

## 3 COLLECTING STYLE SIMILARITY INFORMATION

This section describes the process of collecting data from humans about the style similarity of 3D shapes. Since it is difficult for humans to provide absolute similarity values (for example, to provide a real number to say how stylistically similar a chair model is to a table model), we ask humans to provide relative values. We differentiate between crowdsourced data collection with many users (similar to recent work in this topic [Liu15, Lun15]) and user-guided data collection.

### 3.1 Crowdsourced Data Collection

We collect data by gathering the preferences of a large number of humans by posting tasks on Amazon Mechanical Turk. This idea is similar to previous work [Gar14, Liu15, Lun15] and we describe our process here for completeness. The key is to collect data in the form of triplets where we have three objects (A, B, C) and A is more similar in style to B than C. To collect such triplets, we create queries where a human is presented with a 3D model of one object type X and six models of object type Y. Figure 5 shows some example queries. The task is to identify which two of the six of type Y are more similar in style to the model of type X. For each task, we get eight triplets. If we let the two preferred type Y be $Y_1$ and $Y_2$ and the rest be $Y_3$ to $Y_6$, the eight triplets are of the form (X, $Y_1$, $Y_3$-$Y_6$) and (X, $Y_2$, $Y_3$-$Y_6$).

We generate images of the 3D models such as the one in Figure 5 and post them as HITs (Human Intelligence Tasks) on Mechanical Turk. Each HIT contains 25 tasks and we paid $0.15 for each HIT. We can choose the 3D models in these tasks manually or with an iterative approach (Section 4.2). We generate tasks with various pairs of object types, as indicated in Figure 6.

Each human "Turker" is initially given written instructions and an example task with the responses (two pairs) already chosen by us. We ask Turkers to specifically pay attention to the overall shape, shape of parts, color, and texture before providing their preferences. For the crowdsourced data collection, we had 220 users and collected 48,000 triplets.

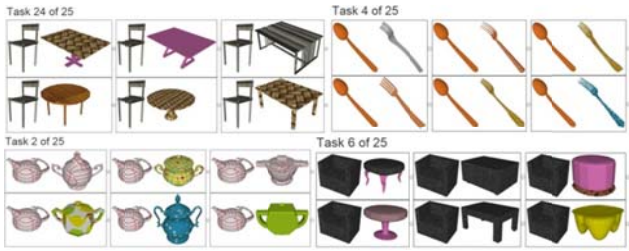

Figure 5: Four example HIT tasks. For each task, users were asked to select two pairs of models out of the six that are more similar in style compared to the others. Users were instructed to compare the following to make their decision: number of parts and their arrangement, color, texture, dimensions of parts and overall shape, curviness of parts and overall shape.

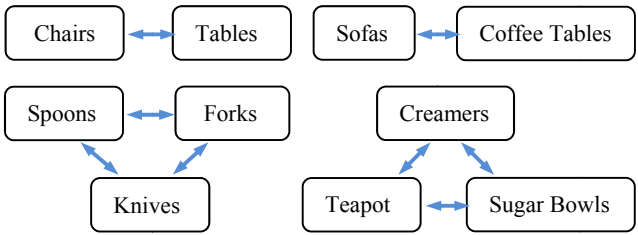

Figure 6: Object types in HIT task. Pairings of 3D model types for which crowdsourcing queries were generated.

### 3.2 User-Guided Data Collection

We also investigate to see if we can learn user-guided metrics and hence we collect data from individuals. We do not use Mechanical Turk here as it can be difficult to require a specific Turker to work on many HITs to collect the needed data (as many Turkers would do just one HIT). Hence we have users who directly use our tools in our lab to collect personalized data. We have two ways to collect such data and we use both of them and combine the data.

We first built a tool to allow users to specify their own preferences by interactively re-ranking search results. The idea is that if a user is not satisfied with the results from the crowdsourced metric, he/she can re-rank the results to generate training data which can then be used to learn a user-guided metric. The tool (Figure 7 top) allows a user to visualize all 3D models of an object type on the left scrollable panel, where the models can be ranked according to their style similarity to a selected 3D model in the current environment on the right. A user is first asked to perform a search with the tool using the crowdsourced metric. The user can then re-arrange the ranked results based on his/her preferences of how well they match in style with a model selected in the current environment. The user is asked to specifically place the ten "closest match" at the top since we use them to generate triplets data. The user interface consists of dragging and dropping the images of the 3D models interactively with the mouse to re-order them. If there is a long list of 3D models in the scrollable panel, the user can also move the mouser cursor over a 3D model and press a key on the keyboard to move it to the top of the ranking. After re-arranging the models, the user clicks a button to generate new triplets according to the ranking. The triplets are of the form (A, B, C) where A is the selected model in the environment, B is one of the top ten ranked models, and C is one of the other models (not ranked as top ten). Such triplets indicate that for the selected model A, the model B is more similar in style to it than C. This process generates 10*(n-10) triplets where n is the number of models we have for the object type being ranked. Hence we choose this method as it is an effective way to generate a large number of triplets.

The user can then use another tool (Figure 7 bottom) to generate more triplets. For this tool, the user can choose two object types X and Y, and the system randomly chooses one model of type X and six models of type Y. The format of these six pairs of models is the same as the HIT task. The user chooses two of the six and the system generates eight corresponding triplets. This task can be repeated as the user wishes to generate triplets.

For this user-guided data collection process, we had five users who collected data for various object types. Each user generated just over 30,000 triplets and took about 45 minutes.

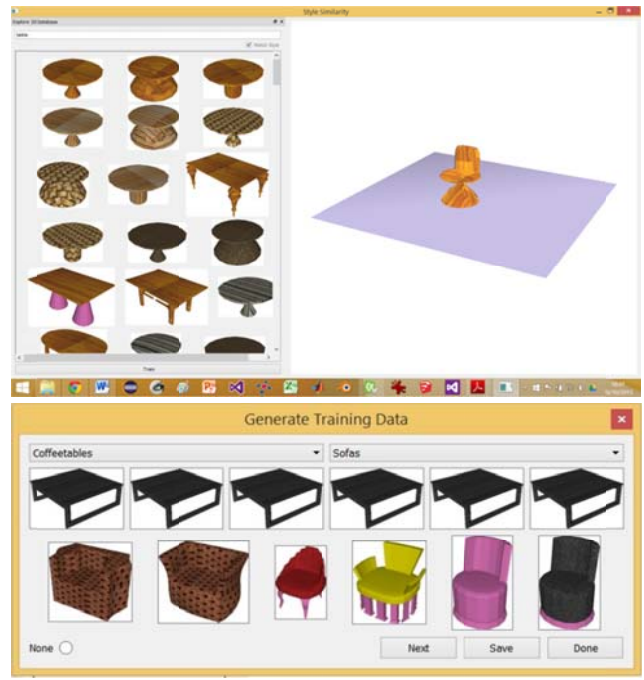

Figure 7: User-Guided Data Collection. Top: A tool for the user to specify style preferences. Right window shows a 3D environment. Left window shows a list of 3D models of an object type. This list can be ranked based on the style similarity compared to the selected model on the right. The user can interactively drag and drop these models to re-rank them to specify their own style preferences, and then the metric can be re-trained. Bottom: Another tool for the user to generate as many additional triplets as he/she wishes. It uses a format that is similar to the HIT tasks.

## 4 LEARNING SIMILARITY METRIC

In this section, we describe our framework for learning a style similarity metric with the feature vector and similarity data described in Sections 2 and 3. The framework is based on metric learning and is inspired by previous methods [Gar14, Sal15]. Our work takes an iterative approach to compute a style metric.

### 4.1 Style Metric Computation

We use a metric learning approach to compute a distance between two 3D models based on their style similarity. Let **x** and **y** be the feature vectors for two 3D models, and we wish to compute the distance between them:

$$d(x, y) = \sqrt{(x - y)^T W (x - y)}$$

The learning formulation and solution to solve for **W** is the same as in previous approaches [Gar14, Sal15], and hence we do not repeat the details here but refer the reader to the previous works.

### 4.2 Iterative Approach

We take an iterative approach to learn a metric and the idea is to gradually build a better **W** matrix. We take the following steps for each pair of object types X and Y:

(1) Initialize **W**$_{current}$ to identity matrix or random matrix
(2) Repeat
   (3) Generate triplets of the form $(X_i, Y_i, Y_j)$,
      where $X_i$ is a random 3D model of type X,
      $Y_i$ is 3D model with feature vector $y_i$
      such that $d(x_i, y_i) \leq d(x_i, y)$ for all Y models
      computed with **W**$_{current}$ in the $d()$ function,
      and $Y_j$ is random 3D model of type Y that is not $Y_i$
   (4) Use above triplets to learn **W**$_{new}$
   (5) Set **W**$_{current}$ = **W**$_{new}$

We post HITs on Amazon Mechanical Turk to collect data and learn the weight matrix in each iteration. We choose to repeat the iteration process until the accuracy improves by less than two percent. We can compute the prediction accuracy of a metric learned with a set of triplets by performing five-fold cross validation on them.

The reliability of Turkers was an issue when collecting crowdsourced data. For each HIT, we manually choose 5 control tasks out of 25 as control questions to check the quality of the responses. We only accept a HIT if 80% or more of the control questions match with our responses. In each iteration, we keep re-posting the rejected HITs until we get the desired number of HITs.

We noticed that the HIT rejection rate tends to be high in the initial iterations (as high as 60% in some cases). This is because the initial iterations produce essentially "random" triplets (i.e. $Y_i$ and $Y_j$ being random due to the initial **W**). Hence it was difficult for Turkers to provide good responses without paying proper attention and many of them gave responses that seem random. As we progress towards more iterations, the learned **W** matrix becomes more effective and the triplets become less "random".

## 5 RESULTS AND APPLICATIONS

We present the results towards each of our four contributions. We use the applications of style similarity based 3D model search and 3D scene composition to demonstrate our work.

**Color and Texture Features.** Figure 8 shows several plots of the learned weight matrices and these are learned with the iterative approach to show the best results. We experimented with both diagonal and full matrices and found no significant difference. Hence we choose to learn diagonal matrices and plot the log of the diagonal values (which correspond to the relative importance of the feature values). The plots show that one of the color-related feature (LBP) consistently dominates over the other color and geometry features, and this is true across the four categories. Hence we suggest the use of this feature (LBP [Oja02]) for future development of these kinds of style similarity metrics. The other pattern we observe in these plots is that there is consistency in the 2600+ feature values again across the four categories, which shows that our method is robust.

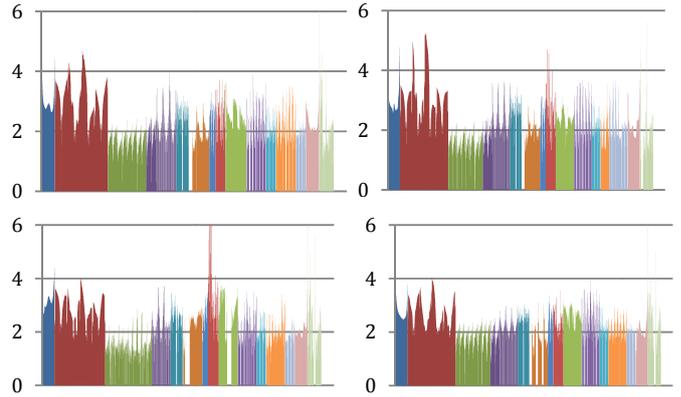

Figure 8: Log plots of the learned weights for (from top left) "dining", "living", "tableware", and "cutlery". The weights correspond to features in the feature vector, in the order described in Section 2. There are 13 geometric features and 5 color/texture features.

Figure 9 shows the results of style similarity based search with our style metric. The top five search results for each query 3D model show that while both geometry and color/texture are important, color/texture is considered first when attempting to match style before geometry is considered. This is true across the different types of shapes that are shown. In the second last row of Figure 9, the red sofa happens to match well with the query model as their curvatures are similar. In Figure 1, we use our style similarity metric with our search tool to compose 3D scenes. As the 3D models that are preferred by the crowdsourced metric are placed at the top of the search results, it is easier to find models that match in style with a selected shape.

Hence we have empirical evidence to support our hypothesis that color/texture features are dominant over geometry features, in the plots of weights and in the style based search results.

**Clustering of Object Types.** Table 1 shows the accuracy results for different pairings of object types and clusters. We do not take the iterative approach here to ensure that the randomness does not affect the results. We instead created HITs manually to cover the range of 3D models in each object type. We took 5 HITs with acceptable responses (after control questions) for each pair of object types, and generated a total of 6040 triplets. For the clustering into groups, the idea is that chairs/tables can be a "furniture" cluster and forks/spoons can be a "cutlery" cluster. We combine the collected triplets from the separate types to create the triplets data for the clusters.

|        | Chairs | Tables | Forks | Spoons |
|--------|--------|--------|-------|--------|
| Chairs |        | 66.73  | 34.52 | 50.44  |
| Tables | 73.29  |        | 41.67 | 47.52  |
| Forks  | 64.25  | 51.19  |       | 80.19  |
| Spoons | 38.87  | 61.17  | 61.38 |        |

|        | Chairs and Tables | Forks and Spoons |
|--------|-------------------|------------------|
| Chairs | 66.73             | 42.24            |
| Tables | 73.29             | 42.99            |
| Forks  | 61.94             | 80.19            |
| Spoons | 50.16             | 61.38            |

|               | Chairs and Tables | Forks and Spoons |
|---------------|-------------------|------------------|
| Chairs Tables | 73.15             | 42.65            |
| Forks Spoons  | 51.83             | 72.21            |

|               | Chairs, Tables, Forks, and Spoons |
|---------------|-----------------------------------|
| Chairs Tables | 56.84                             |
| Forks Spoons  |                                   |

Table 1: Cross-validation percentages for different pairings of object types and clusters. We learn metrics for X→Y, where X (and Y) is the type or cluster in each row (and column).

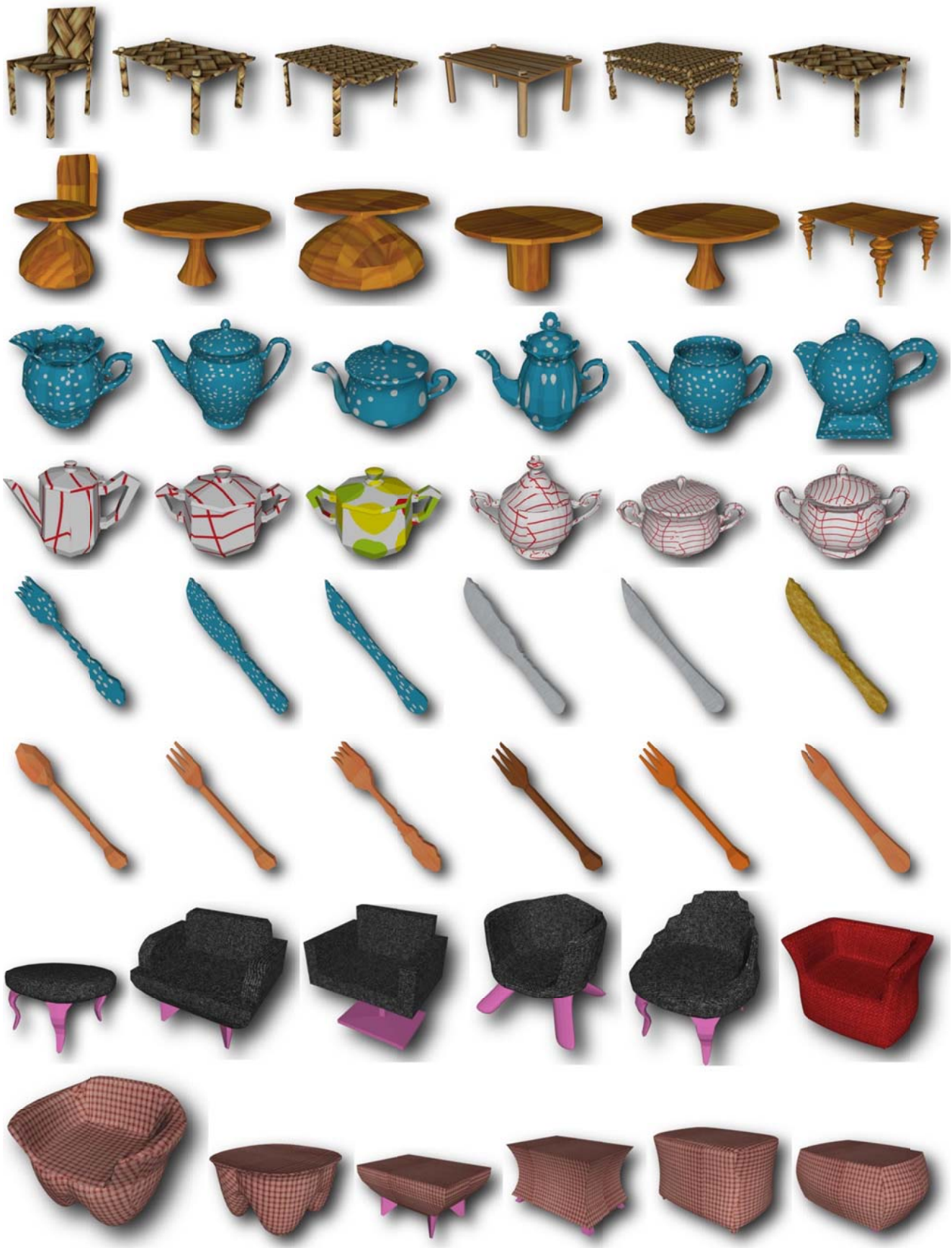

Figure 9: Style similarity based search results with our crowdsourced metric. Each row first shows the query model and then the top five ranked models.

Observing the results from Table 1, we see that the percentages for some object types (e.g. chairs and tables) are comparable to the results with the iterative approach (shown below). We intentionally compared across different object types here (e.g. forks→tables) and hence some pairings give low percentages as it may be difficult to compare between some object types. This does not affect what we aim to show: the tradeoff between learning metrics for specific object types versus clusters of object types.

We observe that the percentages of the clustered pairings are somewhat averaged from the percentages of the separated pairings. We hypothesized that the clustered metrics would be less accurate, as they may be mixing object types that are quite different. However, our empirical results show no clear consensus of whether the metrics from specific object pairings or clustered pairings is better.

Since we combine the triplets data to learn a style metric during this "clustering" process, we also tested whether the number of triplets would have been a variable that affects the percentages (i.e. more triplets data may lead to a higher percentage). Table 2 shows the results where we randomly take half of the triplets in each case and re-calculate the percentage. These results show that the number of triplets does not affect the percentage.

|  | Chairs and Tables | Forks and Spoons |  | Chairs, Tables, Forks, and Spoons |
|---|---|---|---|---|
| Chairs Tables | 72.30 | 40.68 | Chairs Tables | 56.36 |
| Forks Spoons | 52.12 | 71.10 | Forks Spoons | |

Table 2: We randomly take half of the original triplets (compared to Table 1) in each of these five cases and re-calculate the cross-validation percentages.

**User-Guided Style Similarity Metrics.** Figure 10 shows that the user-guided results are interestingly somewhat different from the crowdsourced results. For example, the color and shape of the legs of the furniture pieces are different between the two results. For the "cutlery" example, the crowdsourced results mainly match with the color/texture features, while the user-guided results include a spoon that has very similar geometry (i.e. round-shaped handle) but different color/texture. The individual user in this case was attentive to the geometry of the cutlery in addition to their color/texture. These results demonstrate that we can learn a style metric for individual users that is different from the crowdsourced metric, providing evidence for our hypothesis that the user-guided concept can be beneficial in some cases. However, since the overall color and shape preferences among different people are still mostly consistent, the differences in the user-guided metrics may only be subtle.

**Iterative Learning Approach.** We collected 10 HITs of crowdsourced data in each iteration. The experiments (Figure 11) support our hypothesis that the iterative process can generate useful HITs, and can avoid having to randomly generate triplets [Liu15] or to manually group the 3D models in advance [Lun15]. We found that we can stop after three iterations, and this was consistent across the four object categories.

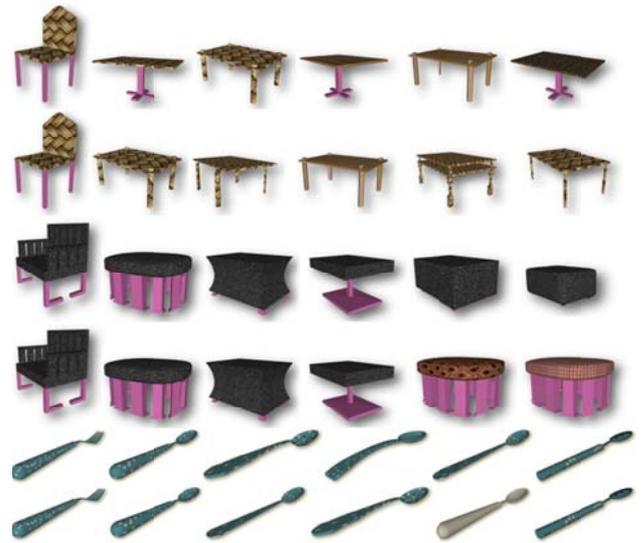

Figure 10: Comparison between crowdsourced (first row in each case) and user-guided (second row in each case) results for "dining", "living", and "cutlery" categories. Each row first shows the query model and then the top five ranked models from style similarity based search.

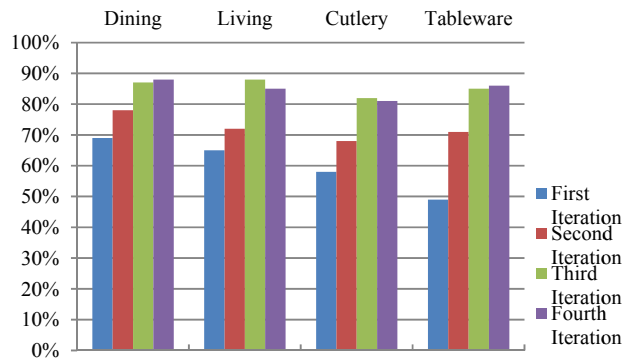

Figure 11: Cross-validation percentages for the iterative learning for "dining" (chairs→tables), "living" (sofas→coffee tables), "cutlery" (spoons→forks), and "tableware" (teapots→sugar bowls).

## 6 Discussion, Limitations, and Future Work

Crowdsourcing and learning approaches have recently been successfully applied to various computer graphics problems. In particular, this approach has been applied to compute a style similarity metric for 3D models. In this paper, we further explore this problem and provide four new contributions.

In addition to the color and texture features, other properties such as construction material (e.g. glass or metal) and how textures are mapped to 3D shapes (not just the texture image) can be taken into consideration in future work when computing features.

The results from our clustering experiments are different from the results presented in Liu et al. [Liu15] although they do not cluster the object types as we do. In their "all triplets" scenario where triplets of different pairs of object types are combined, their results show a better percentage accuracy compared to when the triplets of different objects types are separated trained to form metrics. In the case where we combined the four object types in our experiments, our results show a similar percentage accuracy compared to metrics computed with separate object types. Future work can further investigate the reason for this discrepancy. A larger number of object types and a larger dataset may give more insightful results.

One limitation to the current learning method is that the distance function is a simple Euclidean distance metric. Hence the learned style metrics are limited in their expressive power, and more complex non-linear functions can allow the metrics to better represent human preferences.

### REFERENCES

[1] D. Chen, X. Tian, Y. Shen, and M. Ouhyoung. On Visual Similarity Based 3D Model Retrieval. *Computer Graphics Forum*, 22(3):223-232, 2003.

[2] J. Fogarty, D. Tan, A. Kapoor, and S. Winder. CueFlik: Interactive concept learning in image search. *SIGCHI Conference on Human Factors in Computing Systems*, 29-38, 2008.

[3] T. Funkhouser, M. Kazhdan, P. Min, and P. Shilane. Shape-based Retrieval and Analysis of 3D Models. *Communications of the ACM*, 48(6):58-64, 2005.

[4] E. Garces, A. Agarwala, D. Gutierrez, and A. Hertzmann. A similarity measure for illustration style. *ACM Transactions on Graphics*, 33(4):93, 2014.

[5] A. Kovashka and K. Grauman. Attribute adaptation for personalized image search. *ICCV*, 3432-3439, 2013.

[6] B. Kulis. Metric Learning: A Survey. *Foundations and Trends in Machine Learning*, 5(4):287-364, 2013.

[7] T. Liu, A. Hertzmann, W. Li, and T. Funkhouser. Style Compatibility for 3D Furniture Models. *ACM Transactions on Graphics*, 34(4):85, 2015.

[8] Z. Lun, E. Kalogerakis, and A. Sheffer. Elements of style: learning perceptual shape style similarity. *ACM Transactions on Graphics*, 34(4): 84, 2015.

[9] P. O'Donovan, J. Libeks, A. Agarwala, and A. Hertzmann. Exploratory Font Selection Using Crowdsourced Attributes. *ACM Transactions on Graphics*, 33(4):92, 2014.

[10] T. Ojala, M. Pietikainen, and T. Maenpaa. Multiresolution gray-scale and rotation invariant texture classification with local binary patterns. *IEEE PAMI,* 24(7):971-987, 2002.

[11] R. Osada, T. Funkhouser, B. Chazelle, and D. Dobkin. Matching 3D Models with Shape Distributions. *Shape Modeling International*, 154-166, 2001.

[12] G. Pasi. Issues in Personalizing Information Retrieval. *IEEE Intelligent Informatics Bulletin*, 11(1):3-7, 2010.

[13] B. Saleh, M. Dontcheva, A. Hertzmann, and Z. Liu. Learning style similarity for searching infographics. *Graphics Interface*, 59-64, 2015.

[14] J. Sang, X. Changsheng, and L. Dongyuan. Learn to personalized image search from the photo sharing websites. *IEEE Transactions on Multimedia,* 14(4): 963-974, 2012.

[15] T. Surazhsky, E. Magid, O. Soldea, G. Elber, and E. Rivlin. A comparison of Gaussian and Mean curvatures estimation methods on triangular meshes. *ICRA*, 1021-1026, 2003.

[16] O. Tamuz, C. Lu, S. Blongie, O. Shamir, and A. Kalai. Adaptively Learning the Crowd Kernel. *ICML*, 673-680, 2011.

[17] J. Tangelder and R. Veltkamp. A survey of content based 3D shape retrieval methods. *Multimed Tools Appl*, 39:441-471, 2008.